\documentclass[aps,prb,showpacs,superscriptaddress,groupedaddress,notitlepage,amsmath]{revtex4-1}
\pdfoutput=1
\usepackage[caption=false]{subfig}
\usepackage{natbib}
\usepackage{todonotes}
\usepackage{floatrow} \usepackage{subfig} \usepackage{gensymb}
\usepackage{graphicx}\usepackage{epstopdf}
\usepackage[version=3]{mhchem} \usepackage{dcolumn}\usepackage{bm}\usepackage{placeins}
\usepackage{hyperref}\usepackage{cleveref}

\begin{document}
\author{Anirudh Raju Natarajan}
\email{anirudh@ucsb.edu}
\affiliation{Materials Department, University of California, Santa Barbara, 93106, CA, USA}
\author{Anton Van der Ven}
\email{avdv@ucsb.edu}
\affiliation{Materials Department, University of California, Santa Barbara, 93106, CA, USA}
\title{Linking electronic structure calculations to generalized stacking fault energies in multicomponent alloys}
\begin{abstract}
The generalized stacking fault energy is a key ingredient to mesoscale models of dislocations. Here we develop an approach to quantify the dependence of generalized stacking fault energies on the degree of chemical disorder in multicomponent alloys. We introduce the notion of a \emph{``configurationally-resolved planar fault''} (CRPF) energy and extend the cluster expansion method from alloy theory to express the CRPF as a function of chemical occupation variables of sites surrounding the fault. We apply the approach to explore the composition and temperature dependence of the unstable stacking fault energy (USF) in binary Mo-Nb alloys.
First-principles calculations are used to parameterize a formation energy and CRPF cluster expansion.
Monte Carlo simulations show that the distribution of USF energies is significantly affected by chemical composition and temperature. The formalism can be applied to any multicomponent alloy and will enable the development of rigorous models for deformation mechanisms in high-entropy alloys.
\end{abstract}
\maketitle

\section{Introduction}
The effect of compositional fluctuations and configurational ordering on the properties of a dislocation is a long-standing problem in materials science\cite{hull2011,bulatov2006,laughlin2014}. Experimental and computational studies of complex-concentrated alloys, also referred to as ``\emph{high-entropy alloys}''\cite{yeh2004,cantor2004, george2019}, have revealed that dislocation motion, core structure and stacking fault energies can vary significantly with the local ordering of chemical species\cite{rao2017,rao2017a, ding2019, ding2018, li2019}. Even chemically similar alloys can have drastically different mechanical properties\cite{miracle2016}. Rapid changes in mechanical properties due to minor variations in chemistry undoubtedly have their origin in atomistic deformation mechanisms. Models that are able to link the properties of a dislocation to the degree of long- and short-range chemical ordering in multicomponent alloys are therefore necessary to provide fundamental insights about the role of chemistry on mechanical properties. \par

The generalized stacking fault energy (GSF), commonly referred to as the $\gamma$-surface, plays an important role in quantifying dislocation properties\cite{bulatov2006,vitek1968,vitek1973,vitek1992}.
The GSF energy is equal to the work required to displace two halves of a perfect crystal relative to each other along a particular crystallographic plane.
It is an essential ingredient in Peirls-Nabarro \cite{peierls1940,nabarro1947,lu2000} and phase-field models \cite{shen2003,shen2004,koslowski2002a,hunter2011} of dislocations, where it is used to assess the energy penalty due to a disregistry between the adjacent crystallographic planes across the slip plane of a dislocation.
The GSF energy can also provide qualitative insights about dislocation core structures and preferred partial dislocation structures \cite{vitek2008}. \par

Here we develop a method that rigorously captures the dependence of the GSF energy on the degree of ordering in multicomponent alloys.
We extend the cluster expansion formalism of alloy theory to describe the energy of displacing and cleaving two halves of a crystal relative to each other as a function of descriptors of the degree of chemical order.
We then apply the method to a study of unstable stacking fault energies in the binary Mo-Nb alloy and use Monte Carlo simulations to quantify the average unstable stacking fault energy as a function of temperature and composition.
Our study shows that the GSF energy of the Mo-Nb alloy has both a strong composition and temperature dependence.

\section{Formalism}
\begin{figure}[htbp]
    \centering
    \includegraphics[scale=0.2]{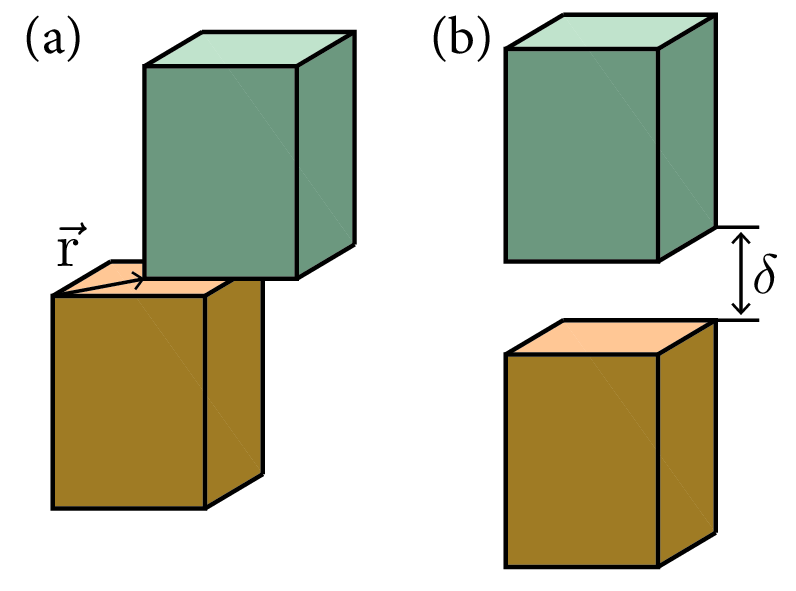}
    \caption{Schematic figure showing the possible ways to shift two rigid solid blocks relative to each other once a particular slip plane has been defined}
    \label{fig:slab_schematic}
\end{figure}

Two parts of a crystal can be shifted relative to each other by a vector $\vec{r}$ that is parallel to a glide plane as shown in \cref{fig:slab_schematic}(a) or the crystal can be cleaved by a distance $\delta$ perpendicular to the glide plane as in \cref{fig:slab_schematic}(b). The gliding of the two parts of the crystal relative to each other results in a planar fault. The energy per unit area as the two halves of a crystal are shifted relative to each other by $\vec{r}$ is conventionally referred to as the GSF energy. It can be defined to be the glide energy either at fixed $\delta$ or at a value of $\delta$ corresponding to zero tractions perpendicular to the glide plane. Throughout we will use periodic boundary conditions. \par

\begin{figure}[htbp]
    \centering
    \includegraphics[scale=0.4]{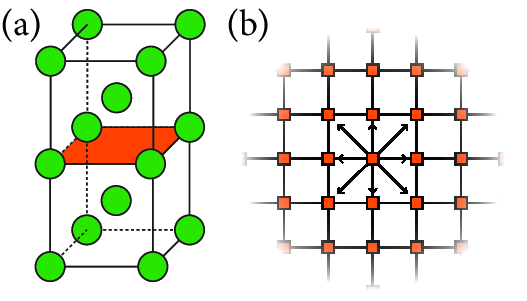}
    \caption{(a) Schematic figure showing the $(001)$ cut-plane on a conventional bcc crystal. (b) The two-dimensional glide vector space formed by the relative translations that recover the perfect bulk crystal. The energy of a pure bi-crystal is periodic, with the same energy being recovered at glide vectors shown with squares in the figure.}
    \label{fig:lattice_schematic}
\end{figure}
The GSF energy of a single component crystal is a periodic function over the space of two-dimensional glide vectors $\vec{r}$. Translating the two halves of a crystal by a full lattice vector recovers the unfaulted bulk crystal. Each point in the two-dimensional glide space $\vec{r}$ that coincides with a translation vector will therefore have the same GSF energy. Figure \ref{fig:lattice_schematic}(a) shows a $(001)$ glide plane of a body-centered cubic (bcc) crystal. The corresponding two-dimensional glide vector space for the $(001)$ glide plane of bcc is shown in \cref{fig:lattice_schematic}b with glide vectors that recover the bulk bcc crystal shown as squares. Glide vectors that differ from a lattice translation produce a bicrystal with a planar fault. The introduction of a planar fault usually increases the energy of the crystal. \par

\begin{figure}[htbp]
    \centering
    \includegraphics[scale=0.4]{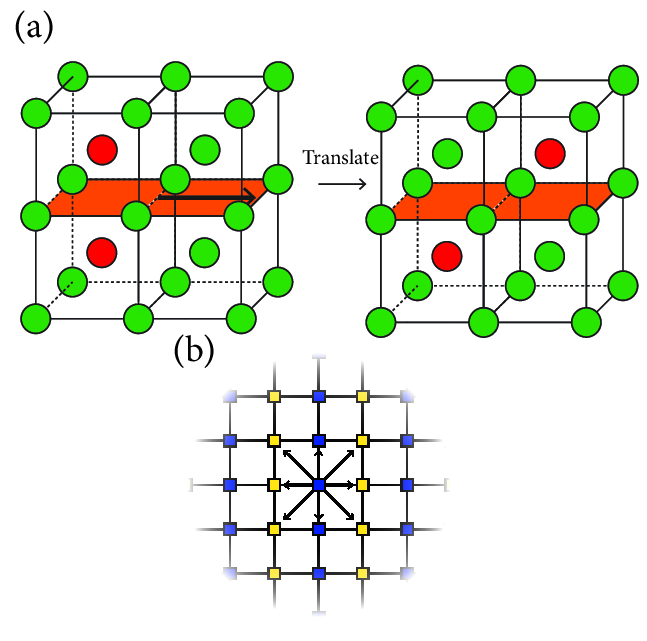}
    \caption{(a) Schematic figure showing a periodic ordering on bcc. Upon translating the top and bottom parts of the crystal by a conventional bcc lattice vector (indicated by the thick black arrow), a different local ordering arises in the vicinity of the cut-plane. The planar defect is akin to the well-known ``anti-phase'' boundaries in metallurgy. (b) The two-dimensional glide vector space for the ordering shown in (a). Squares with the same color correspond to the same ordering and hence the same energy.}
    \label{fig:apb_bcc_shift}
\end{figure}

The GSF energies of multicomponent alloys differ from those of pure elements. The periodicity of the GSF energy that characterize a single component crystal is generally broken in a multicomponent alloy. This is illustrated for a particular ordering of red and green atoms on bcc in \cref{fig:apb_bcc_shift}(a). A glide of the upper half of the crystal relative to the lower half by a $[100]$ translation vector of the underlying bcc crystal structure results in a different ordering of red and green atoms. Though the bcc crystal structure is recovered, a planar defect referred to as an anti-phase boundary has been created. The energy of the crystal before and after the glide by a bcc translation vector is therefore no longer the same since the arrangement of red and green atoms has changed. Two lattice translations along the $(001)$ plane are required for the example of \cref{fig:apb_bcc_shift}(a) to recover the original ordering. Figure \ref{fig:apb_bcc_shift}(b) shows the symmetry in the two-dimensional space of glide vectors $\vec{r}$, with purple squares corresponding to the original ordering and yellow squares corresponding to orderings with an anti-phase boundary. For the particular ordering in \cref{fig:apb_bcc_shift}(a), glide vectors along the $[010]$ direction do not change the ordering and the energy remains unchanged. \par

\begin{figure}
    \centering
    \includegraphics[scale=0.4]{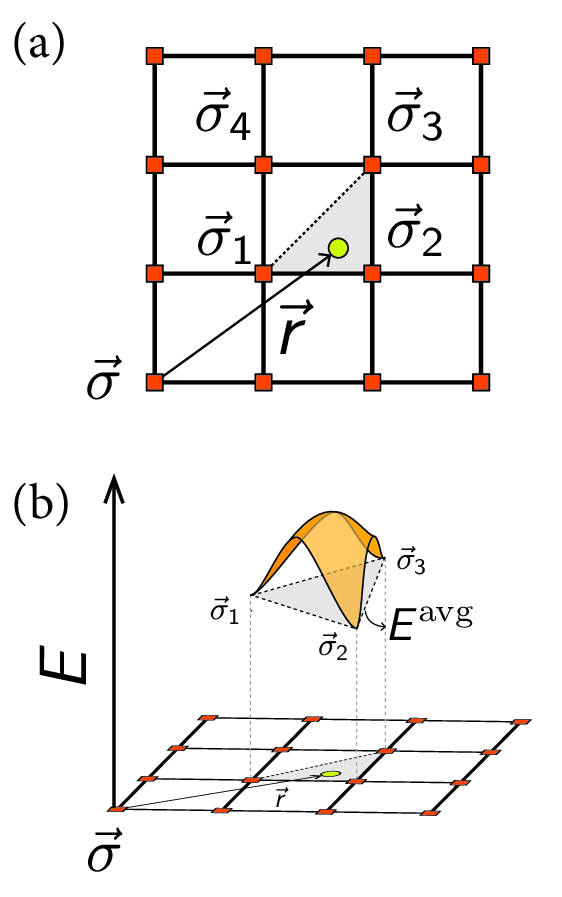}
    \caption{(a) Schematic two-dimensional glide vector space for a bicrystal, with four different orderings arising at full lattice translations, corresponding to squares. (b) Schematic energy of a bicrystal being sheared to transform between three orderings labeled $\sigma_{1}$, $\sigma_{2}$ and $\sigma_{3}$. The bulk configurational contribution is labeled $E^{\textrm{avg}}$.}
    \label{fig:sra_schematic}
\end{figure}

The example of \cref{fig:apb_bcc_shift}(a) shows that it is necessary to track the degree of order in a multicomponent alloy since glides by a translation vector of the underlying parent crystal structure can change the arrangement of chemical species in the crystal. The state of ordering in a multicomponent solid can be described mathematically by assigning occupation variables to each site of its crystal. Consider a large crystal with $N$ sites (assuming periodic boundary conditions) where each site can be occupied by one of two chemical components A or B. Any ordering of A and B atoms on this crystal can be represented with an occupation vector $\vec{\sigma} = \{ \sigma_{1}, \sigma_{2}, \cdots ,\sigma_{N} \}$, where $\sigma_{i}$ is an occupation variable that takes the value of +1 if site $i$ is occupied by A and a value of -1 otherwise. The labels $i$ refer to sites in a reference crystal that is not deformed. For the purpose of tracking the configuration of a deformed crystal, we map each site of the deformed crystal onto the nearest site of the reference crystal. A particular configuration of A and B atoms, $\vec{\sigma}_{1}$, for example, may then be converted into a new configuration $\vec{\sigma}_{2}$ upon application of a glide that coincides with an elementary translation vector of the underlying parent crystal structure. A second glide by another translation vector may convert $\vec{\sigma}_{2}$ into $\vec{\sigma}_3$. The changes in configuration upon application of glides coinciding with parent crystal translations can be represented in the two-dimensional glide space of $\vec{r}$ as schematically illustrated in \cref{fig:sra_schematic}(a). Since the energy of the crystal depends on how the A and B atoms are arranged, it will also vary upon the application of a glide that is equal to a parent crystal translation. This is schematically illustrated in \cref{fig:sra_schematic}(b). \par

While the gliding of two halves of a crystal by a translation vector of the parent crystal changes the configuration of the alloy and therefore its energy, most of the change in ordering is restricted to the vicinity of the glide plane.
The local arrangements of A and B atoms far away from the glide plane are unaffected by the glide since those regions have simply been translated rigidly.
Chemical interactions in an alloy typically decay over a distance of several nanometers when maintaining the solid in a constant state of strain.
The contribution to the energy of the crystal due to a particular arrangement of A and B atoms far away from the glide plane will be identical in two configurations, $\vec{\sigma}_{1}$ and $\vec{\sigma}_{2}$, related by a glide since the local degree of ordering at those large distances from the glide plane are identical in both configurations.
It is only in regions within the chemical interaction range of the glide plane where the local degree of order is different that an energy difference arises.
This motivates the separation of the GSF energy into an average configurational energy and a ``\emph{configurationally-resolved planar fault energy}'' (CRPF) that is a local excess energy.
The energy, $E$, of a bi-crystal (shown schematically in \cref{fig:sra_schematic}(a)) with an initial ordering $\vec{\sigma}$ that is translated by $\vec{r}$ and separated by a distance $\delta$ can then be written as
\begin{equation}
    \label{eq:gsf}
    E(\vec{r},\delta,\vec{\sigma}) = E^{\textrm{CRPF}}(\vec{r}-\vec{r}_{1},\delta,\vec{\sigma}_{1}) + E^{\textrm{avg}}(\vec{r},\vec{\sigma}_{1},\vec{\sigma}_{2},\vec{\sigma}_{3}).
\end{equation}
In this expression, $E^{\textrm{CRPF}}(\vec{r}-\vec{r}_{1},\delta,\vec{\sigma}_{1})$ is the \emph{configurationally-resolved planar fault energy}, with $\sigma_{1}$ being one of the three nearest orderings on the perfect crystal for the glide vector $\vec{r}$ as schematically illustrated in \cref{fig:sra_schematic}(a). The glide vector $\vec{r}_{1}$ converts $\vec{\sigma}$ to $\vec{\sigma}_{1}$. The average configurational energy, $E^{\textrm{avg}}$, is related to the energy of the three nearest configurations as schematically shown in \cref{fig:sra_schematic}(a),  and is defined as
\begin{equation}
\label{eq:avg_config_energy}
    E^{\textrm{avg}}(\vec{r},\vec{\sigma}_{1},\vec{\sigma}_{2},\vec{\sigma}_{3}) = (1-w_{2}-w_{3})E(\vec{\sigma_{1}}) + w_{2}E(\vec{\sigma_{2}}) + w_{3}E(\vec{\sigma_{3}})
\end{equation}
where $E(\vec{\sigma}_{1})$, $E(\vec{\sigma}_{2})$ and $E(\vec{\sigma}_{3})$ are the energies of the $\vec{\sigma}_{1}$, $\vec{\sigma}_{2}$ and $\vec{\sigma}_{3}$ orderings in the perfect crystal. The weights $w_{2}$ and $w_{3}$ are related to the glide vectors by:
\begin{equation}
\label{eq:avg_energy_weight}
    \begin{bmatrix}
        w_{2} & w_{3}
    \end{bmatrix} = \begin{bmatrix}
        \vec{r}_{12}\\
        \vec{r}_{13}
    \end{bmatrix}^{-1} (\vec{r} - \vec{r}_{1})
\end{equation}
where $\vec{r}_{12}$ is the glide vector relating configurations $\vec{\sigma}_{1}$ and $\vec{\sigma}_{2}$ and $\vec{r}_{13}$ connects $\vec{\sigma}_{1}$ to $\vec{\sigma}_{3}$.\par

Rigorous statistical mechanics calculations of the temperature and composition dependence of the GSF energy require the evaluation of $E(\vec{r},\delta,\vec{\sigma})$ across all possible decorations of the bi-crystal. This can be computationally intractable when using quantum mechanical techniques. Surrogate models informed from a small set of quantum mechanical calculations that accurately reproduce the bulk and CRPF energies for arbitrary configurations are thus needed to bridge the gap. In the rest of this section we review the cluster expansion formalism to describe the configurational energy of crystalline solids and subsequently extend it to describe the CRPF energy as a function of configurational ordering.\par

As shown by Sanchez \emph{et al.}\cite{sanchez1984}, the configurational energy $E(\vec{\sigma})$ in a multicomponent solid with a particular crystal structure can be expanded in terms of cluster basis functions according to
\begin{equation}
    \label{eq:clex}
    E(\vec{\sigma}) = V_{0} + \sum_{\alpha} V_{\alpha} \phi_{\alpha}(\vec{\sigma})
\end{equation}
where $E(\vec{\sigma})$ is the energy of $\vec{\sigma}$, $V_{\alpha}$ are expansion coefficients, referred to as effective cluster interactions (ECI), and the $\phi_{\alpha}(\vec{\sigma})$ are cluster basis functions.
For a binary alloy, the cluster basis functions are defined as
\begin{equation}
    \label{eq:cluster_basis_functions}
    \phi_{\alpha} (\vec{\sigma}) = \prod_{j\in \alpha} \sigma_{j}.
\end{equation}
where $\alpha$ refers to a cluster of sites in the crystal, such as pair clusters, triplet clusters etc.
The $V_{\alpha}$ are determined by the chemistry of the alloy.
In most alloys, chemical interactions decay beyond a maximum length and cluster size and the cluster expansion of Eq. \ref{eq:clex} can be truncated.
Clusters related to each other through a symmetry operation in the undecorated crystal have the same expansion coefficient.
Strategies that rely on genetic algorithms\cite{hart2005}, cross-validation\cite{vandewalle2002b}, bayesian regression\cite{mueller2009}, neural networks\cite{natarajan2018} and quadratic programming\cite{huang2017} have been succesfully applied to generate high-fidelity cluster expansion models trained to first-principles calculations.
The resulting lattice models are typically used in conjunction with statistical mechanics tools such as Monte-Carlo simulations to calculate temperature and composition dependent thermodynamic properties of multi-component solids.

The cluster expansion of Eq. \ref{eq:clex} is only valid for a fixed parent crystal structure.
In the context of GSF energy surfaces, it can only be used to describe the energy of the bicrystal for glide vectors $\vec{r}$ that recover the underlying parent crystal structure.
This includes the energies of $E(\vec{\sigma}_{1})$, $E(\vec{\sigma}_{2})$ and $E(\vec{\sigma}_{3})$ appearing in the expression of average configurational energy $E^{\textrm{avg}}$ as defined by \cref{eq:avg_config_energy,eq:avg_energy_weight} and appearing in \cref{eq:gsf}.
We next extend the cluster expansion approach to describe the CRPF energy of \cref{eq:gsf}.\par

We first simplify the problem by exploiting well-established analytical expressions of energy-versus-separation curves to describe the dependence of the CRPF on $\delta$.
For most metals, the energy versus separation curve can be accurately represented with the universal binding energy relation (UBER) of Rose \emph{et al.}\cite{rose1981, enrique2017a} according to
\begin{equation}
    \label{eq:crpf_uber}
    E^{\textrm{CRPF}} = E^{\textrm{CRPF}}_{0} - 2 \kappa \left[-1+ \left(1+\frac{\delta - \delta_{0}}{\lambda}\right)\exp\left(-\frac{\delta - \delta_{0}}{\lambda}\right) \right ]
\end{equation}
where $E^{\textrm{CRPF}}_{0}$ is the CRPF energy at the equilibrium separation $\delta_{0}$, $2\kappa$ is the surface energy at infinite separation, and $\lambda$ is related to the curvature of the energy around the equilibrium separation. The parameters $E^{\textrm{CRPF}}_{0}, \delta_{0}, \kappa, $ and $\lambda$ are all functions of the configuration $\vec{\sigma}$ and glide vector $\vec{r}$.
While Eq. \ref{eq:crpf_uber} is that for the UBER curve, alternate functional forms such as xUBER\cite{enrique2014, enrique2017} may also be employed.\par

The dependence of $E^{\textrm{CRPF}}_{0}$, $\kappa$, $\delta_0$ and $\lambda$ on configuration $\vec{\sigma}$ can be expressed as a cluster expansion. For example, $E^{\textrm{CRPF}}_{0}$ can be written as
\begin{equation}
    \label{eq:local_clex_crpf}
    E^{\textrm{CRPF}}_{0}(\vec{r},\vec{\sigma}) = \Gamma_{0}(\vec{r}) + \sum_{\alpha} \Gamma_{\alpha}(\vec{r}) \phi_{\alpha}(\vec{\sigma})
\end{equation}
where the cluster basis functions, $\phi_{\alpha}$ are the same as those defined in \cref{eq:cluster_basis_functions}, and where $\Gamma_{\alpha}$ are expansion coefficients that are functions of the glide vector, $\vec{r}$. Similar to the cluster expansion of \cref{eq:clex}, the expansion coefficients in \cref{eq:local_clex_crpf} obey certain symmetry properties dictated by the space group of the undecorated bicrystal having undergone a glide $\vec{r}$. Since a glide of a bicrystal by $\vec{r}$ in general breaks symmetry, far fewer expansion coefficients will be equivalent by symmetry than for the cluster expansion of the undeformed parent crystal. For example, translation symmetry in directions perpendicular to the glide plane are lost upon application of a glide $\vec{r}$. Point clusters that are otherwise equivalent by symmetry in the perfect crystal, are no longer equivalent if they are at different distances from the glide plane. The same holds true for multi-body clusters.

The cluster expansions of the parameters $E^{\textrm{CRPF}}_{0}$, $\kappa$, $\delta_0$ and $\lambda$ appearing in Eq. \ref{eq:crpf_uber} extend over all clusters of the bicrystal.
However, these cluster expansions should converge rapidly and only clusters within the chemical interaction range from the glide plane are likely necessary in a truncated cluster expansion.
This becomes evident when rearranging Eq. \ref{eq:gsf} to isolated the CRPF energy according to
\begin{equation}
    \label{eq:crpf_energy}
    E^{\textrm{CRPF}}(\vec{r}-\vec{r}_{1},\delta,\vec{\sigma}_{1}) = E(\vec{r},\delta,\vec{\sigma}) - E^{\textrm{avg}}(\vec{r},\vec{\sigma}_{1},\vec{\sigma}_{2},\vec{\sigma}_{3}).
\end{equation}
The above equation shows that the contribution to the energy of the bi-crystal from regions far away from the glide plane is removed when subtracting off the weighted average energy $E^{\textrm{avg}}(\vec{r},\vec{\sigma}_{1},\vec{\sigma}_{2},\vec{\sigma}_{3})$, since the configurations $\sigma_1$, $\sigma_2$ and $\sigma_3$ have chemical orderings that are identical (up to a translation vector) to that of the bicrystal beyond the chemical interaction range of the glide plane.

In summary, the parameterization of surrogate models that accurately describe the GSF energies in multicomponent alloys requires two separate cluster expansions. The first is a cluster expansion of the formation energies of orderings over the parent crystal structure. This cluster expansion is required to calculate the average configurational energy, $E^{\textrm{avg}}$, in \cref{eq:gsf}. Methods to parameterize these models are well-established. A second cluster expansion is necessary to describe the short-range CRPF energy. A training dataset can be generated by first calculating the bi-crystal energies, $E(\vec{r},\delta,\vec{\sigma})$,  for several symmetrically distinct orderings, $\sigma$, glide vectors, $\vec{r}$, and separation distances, $\delta$. The CRPF energies for each of these configurations can then be calculated with \cref{eq:crpf_energy}. The resulting CRPF energies for a fixed chemical ordering $\sigma$ then serve to train the adjustable parameters of \cref{eq:crpf_uber}, which can then be cluster expanded according to \cref{eq:local_clex_crpf} to describe their dependence on the degree of chemical order.
In most Peirls-Nabarro-type models the bi-crystal is assumed to be under zero stress in the direction perpendicular to the fault. As a result the GSF energy must be minimized relative to the slab separation distance. Thus, the energy given by $E^{\textrm{CRPF}}_{0}$ in \cref{eq:crpf_uber,eq:local_clex_crpf} is the desired quantity when modeling dislocation properties with PN models that assume zero tractions perpendicular to the glide plane. \par

\section{Results}

In this section, we illustrate the above cluster expansion formalism by investigating the composition and configuration dependence of unstable stacking fault energies in the binary Mo-Nb alloy.
The Mo-Nb alloy adopts the bcc crystal structure at all compositions and forms a disordered solid solution between room temperature and the melt \cite{okamoto1991a,jiang2004,blum2005}.
The Mo-Nb binary is of current interest since Mo and Nb are both components of important bcc based high entropy alloys \cite{senkov2018,senkov2019}.
Studies of the mechanical properties of Mo-Nb alloys indicate that their strength is primarily controlled by the formation and motion of screw dislocations and to some extent edge dislocations\cite{maresca2020}.
The screw dislocations are formed along the $\langle 111\rangle$ direction, and are known to spread on the $\{110\}$ planes\cite{vitek2008}.
Much of the physics of dislocations in Mo-Nb alloys is therefore directly related to the unstable stacking fault energy in the $\{110\}$ plane for a relative displacement along the $\langle111\rangle$ direction.

\begin{figure}[htbp]
 \centering
 \subfloat[\label{fig:monb_formation_energy}]{\includegraphics[scale=0.1]{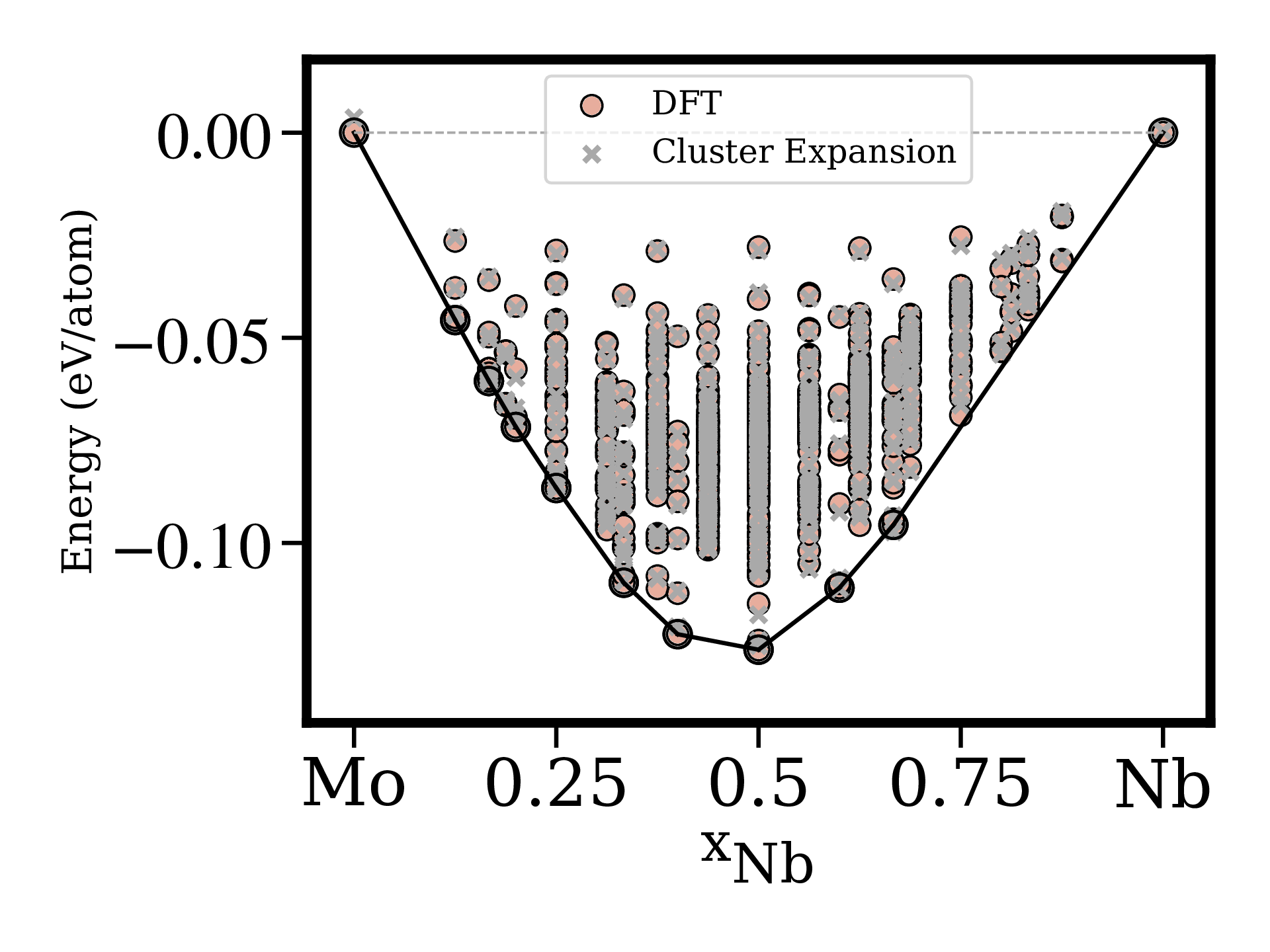}}~
 \subfloat[\label{fig:monb_vol}]{\includegraphics[scale=0.1]{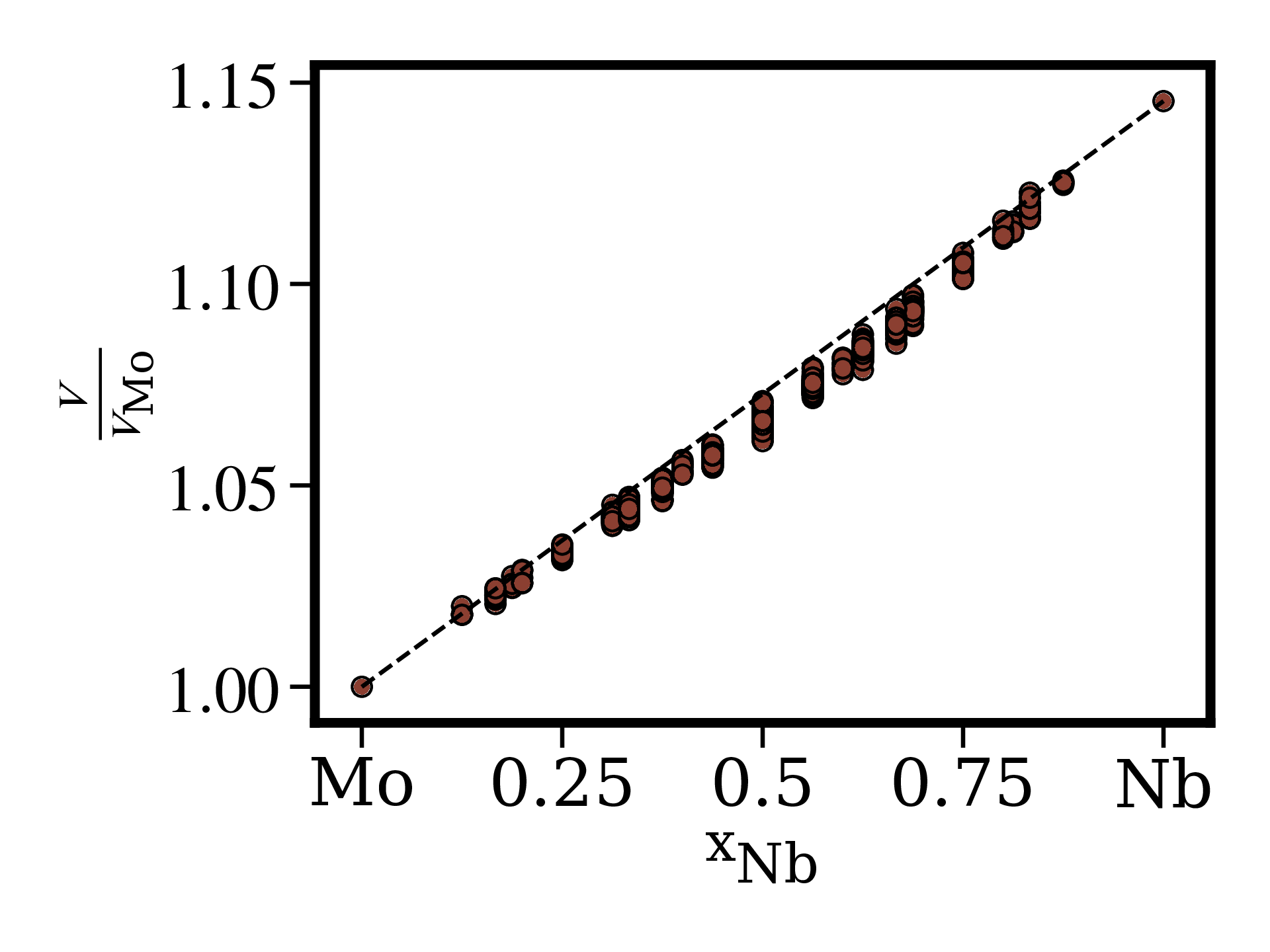}}
 \caption{(a)Comparison of the cluster expanded energies and formation energies calculated from DFT in the binary Mo-Nb alloy on the bcc crystal structure (b) Relaxed volumes of orderings on bcc in the Mo-Nb binary alloy relative to the volume of pure Mo bcc}
\end{figure}
The first step in describing the dependence of the GSF energy on ordering is to construct a cluster expansion for the formation energy of the binary bcc Mo-Nb alloy.
\Cref{fig:monb_formation_energy} shows the formation energies of 847 symmetrically-distinct orderings on the bcc crystal structure in the binary Mo-Nb alloy as calculated with density functional theory (DFT).
The formation energies are referenced to bcc Mo and Nb at 0K.
More details about the DFT calculations and the cluster expansion that was subsequently trained to these energies can be found in \cref{sec:methods}.
The convex hull is outlined in black and shows that several ordered phases are predicted to be stable at 0K.
The energies of the 847 orderings as predicted with a cluster expansion are also shown in \cref{fig:monb_formation_energy} as grey crosses.
The exceptionally low training error of 0.0008 eV/atom and the excellent qualitative agreement between the ground states as predicted with the cluster expansion and those found with DFT suggest that the configurational energy of the Mo-Nb binary alloy is well-described with a truncated cluster expansion model.
\Cref{fig:monb_vol} also shows the relaxed volume of all orderings relative to that of bcc molybdenum. We find them to vary almost linearly as a function of niobium composition, albeit with a slight negative deviation in close agreement with Vegards law\cite{vegard1921} (shown schematically by the dashed line in the figure). \par

\begin{figure*}[htbp]
 \centering
 \subfloat[\label{fig:mo_linescan}]{\includegraphics[scale=0.1]{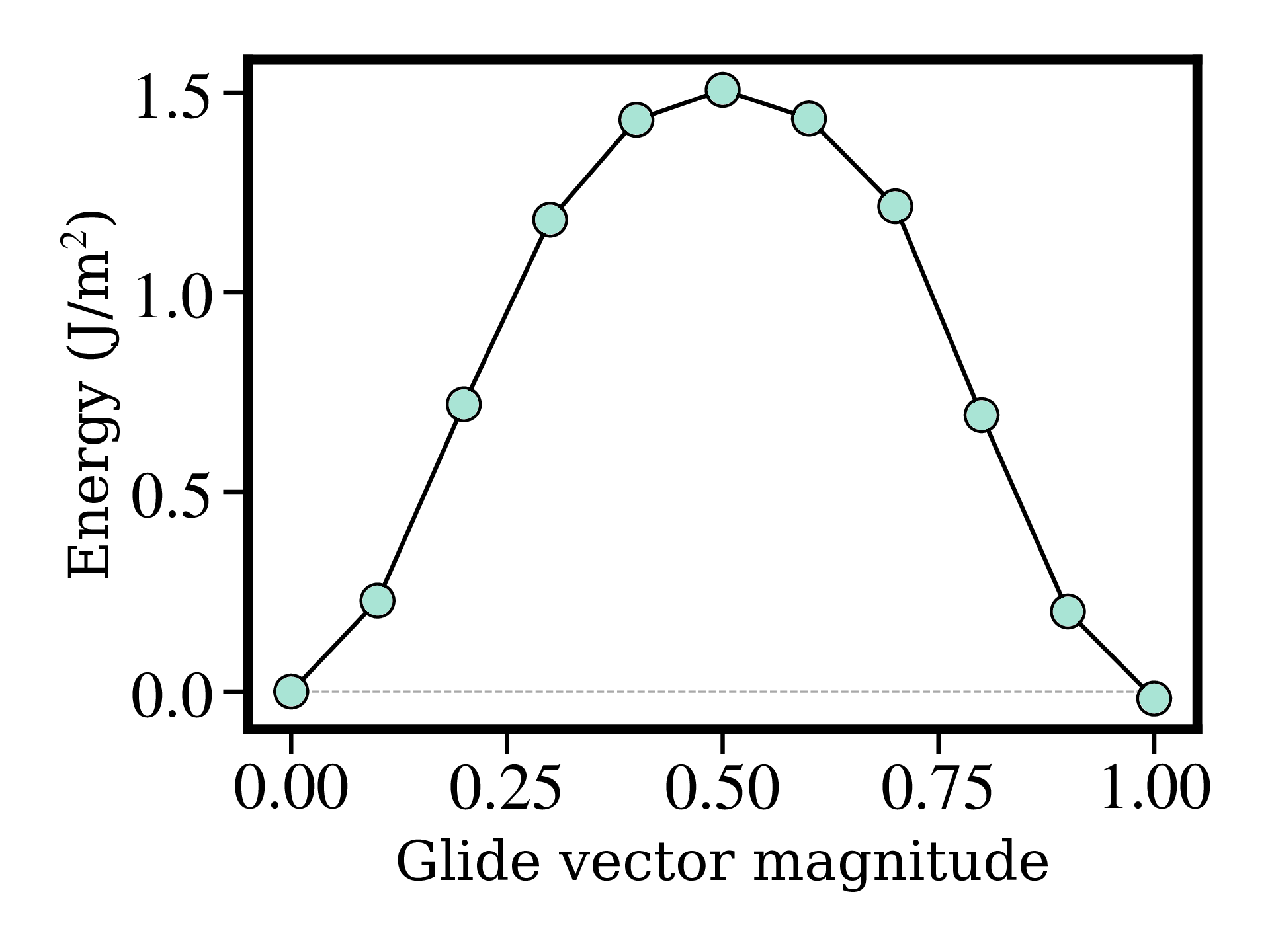}}
 \subfloat[\label{fig:mo_traction_curve}]{\includegraphics[scale=0.1]{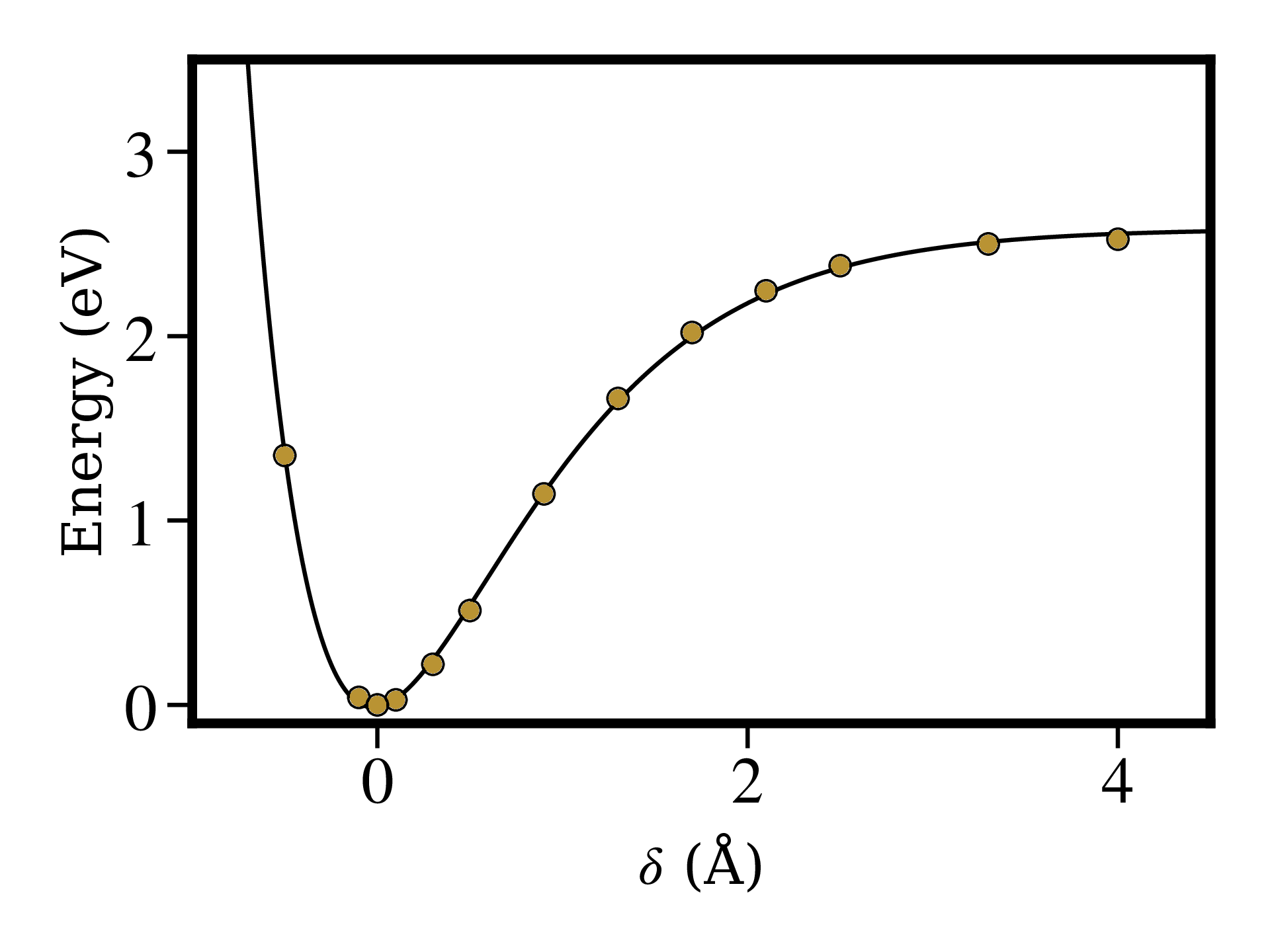}}\\
 \subfloat[\label{fig:B2_linescan}]{\includegraphics[scale=0.1]{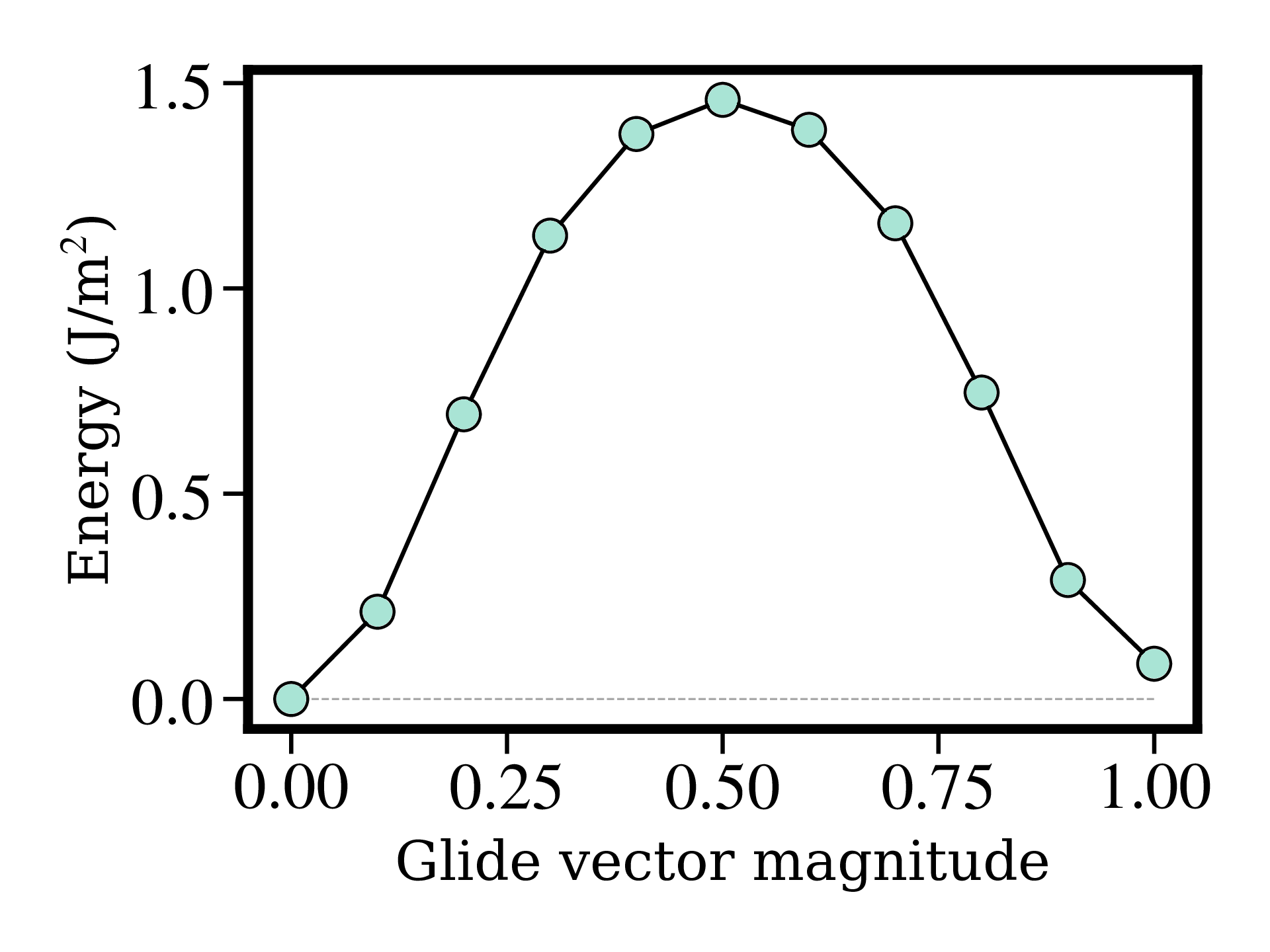}}
 \subfloat[\label{fig:sfe_ordered_phases}]{\includegraphics[scale=0.1]{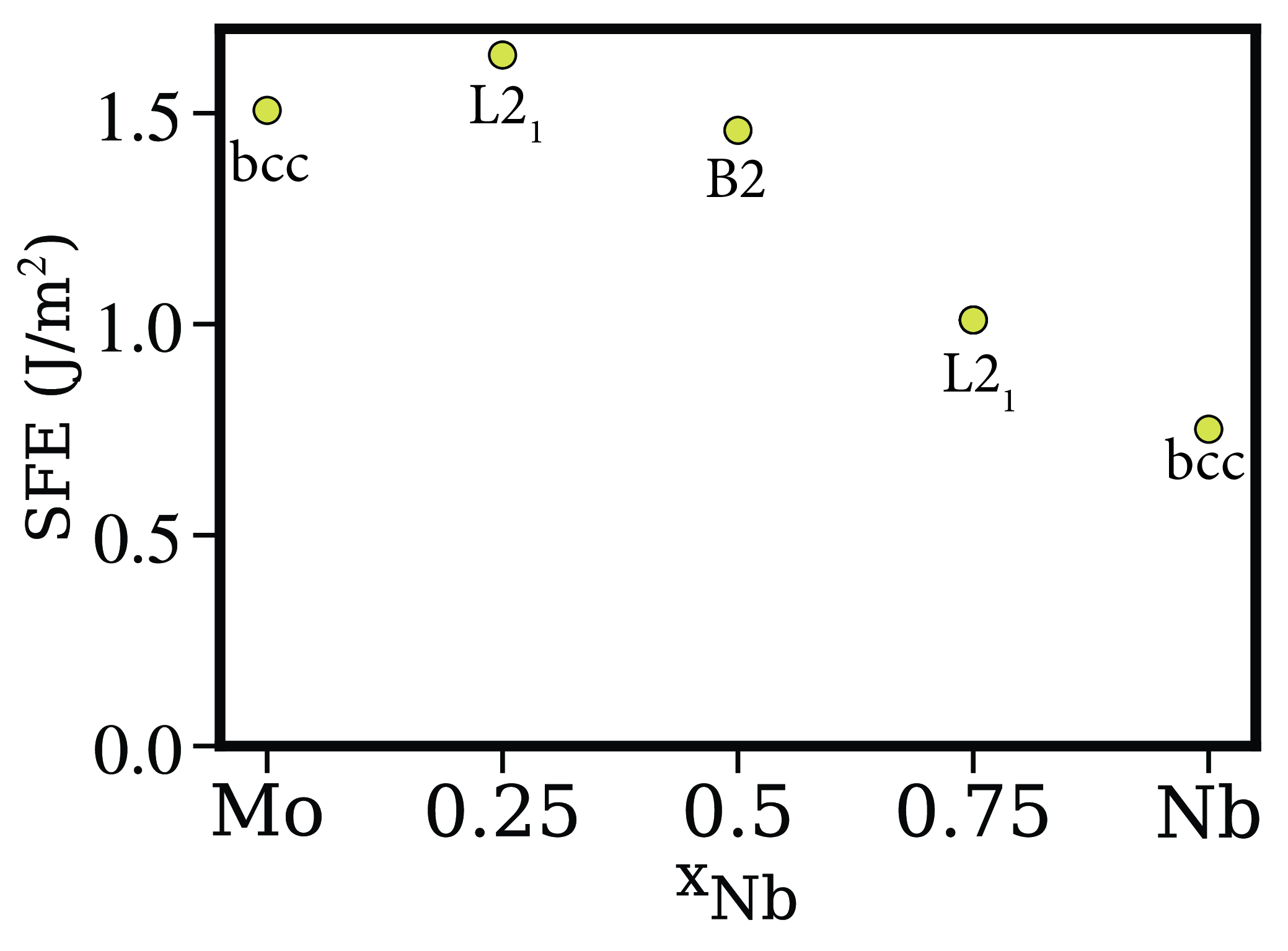}}
 \caption{(a) The GSF energy for a bcc Mo bicrystal translated on the $(110)$ plane along the $[1\overline{1}1]$ direction. (b) The decohesion curves for a pure molybdenum crystal, cleaved along the $(110)$ plane. The energy fit to the UBER is also shown in the dark black line. (c) The GSF energy for a bcc B2 bicrystal translated on the $(110)$ plane along the $[1\overline{1}1]$ direction. (d) The unstable stacking fault energies across the composition range, for an undecorated bcc crystal ($x_{\textrm{Nb}}=0,1$), B2 ($x_{\textrm{Nb}}=0.5$) and L2$_{1}$ ($x_{\textrm{Nb}}=0.25,0.75$) ordering.}
\end{figure*}

The GSF energy surface of a Mo bcc bicrystal for the $\{110\}$ glide plane along the $\langle111\rangle$ direction is shown in \cref{fig:mo_linescan}.
The introduction of a planar fault due to a glide results in an energy penalty.
The fault energy increases until it reaches a maximum at a glide of 1/4$\langle111\rangle$.
This energy corresponds to the unstable stacking fault energy.
As the glide vector approaches a full lattice translation in the $\{110\}$ glide plane (corresponding to 1/2$\langle111\rangle$), the energy decreases until long-range bcc order is restored where the energy becomes equal to that of bcc Mo.
In calculating the GSF as a function of the glide vector $\vec{r}$ of \cref{fig:mo_linescan}, we first calculated the energy of the bicrystal as a function of $\delta$ along the $[110]$ direction to generate decohesion curves for each value of $\vec{r}$.
A particular example of such a curve is shown in \cref{fig:mo_traction_curve}.
The DFT energies as a function of $\delta$ (for fixed $\vec{r}$) were then fit to the universal binding energy relationship (UBER)\cite{rose1981}, and the minimum of each curve was used to construct the GSF of \cref{fig:mo_linescan}.\par

A section of the GSF energy for the B2 ordering with a composition of $x_{\textrm{Nb}}=0.5$ is shown in \cref{fig:B2_linescan}.
Shifting the two halves of a B2 bicrystal through a full lattice translation results in an anti-phase boundary, which is accompanied by an increase of the energy. Similar to pure molybdenum, an unstable stacking fault is found to exist for a glide corresponding to half a translation vector. \Cref{fig:sfe_ordered_phases} collects the calculated unstable stacking fault energies for pure Mo, L2$_{1}$ (Mo$_3$Nb, MoNb$_3$), B2 (MoNb) and pure Nb. We find a strong composition dependence of the unstable stacking fault energies, with the values changing by almost a factor of two with increasing niobium composition. Furthermore, the unstable stacking fault energies vary non-linearly with composition, suggesting that short and long-range order also plays a role in addition to the average concentration. \Cref{fig:sfe_ordered_phases} shows the energy at equilibrium separation for the sheared bicrystal. \par

\begin{figure}[htbp]
    \centering
    \includegraphics[scale=0.1]{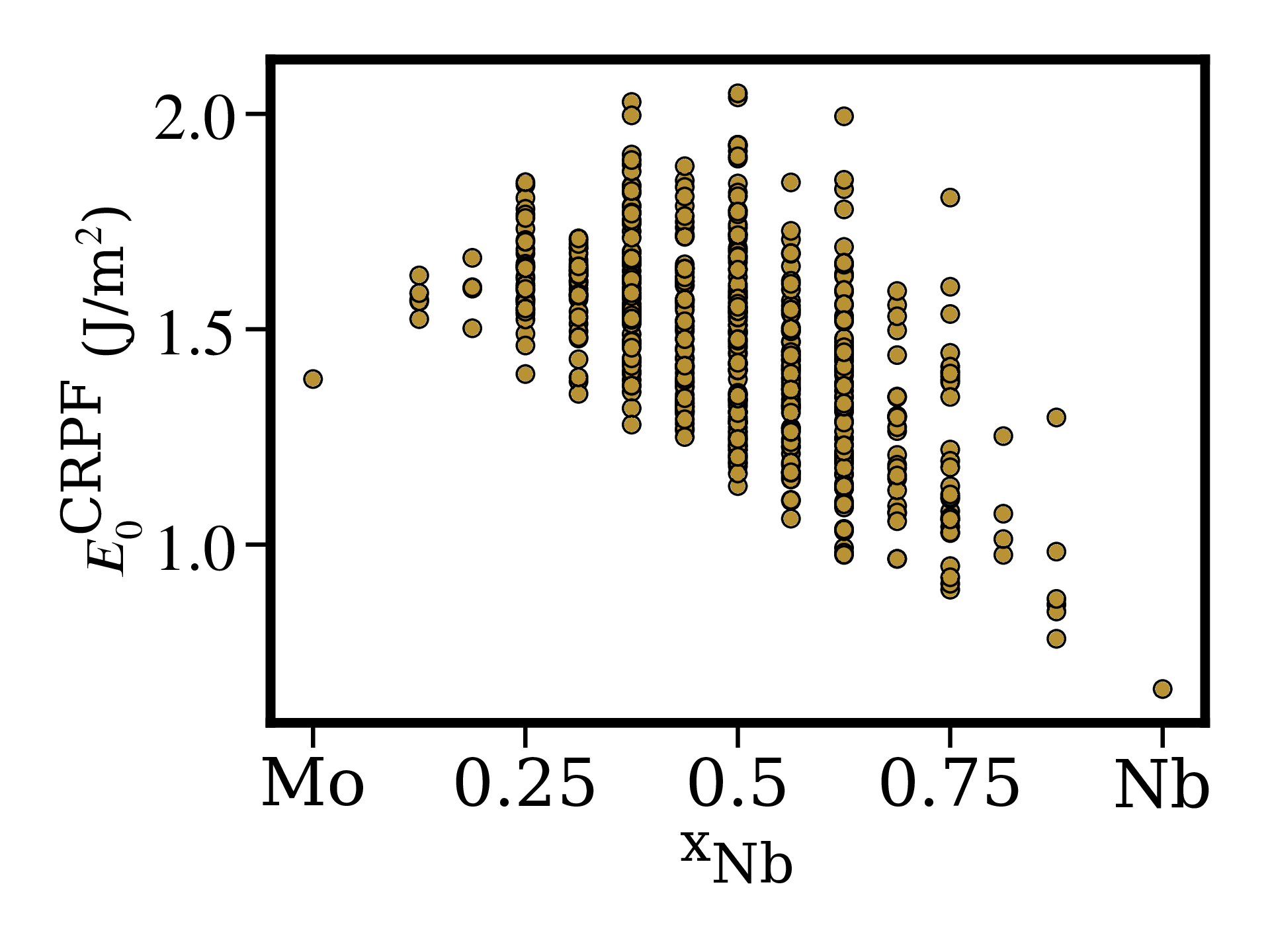}
    \caption{The configurationally-resolved planar fault (CRPF) energy in the Mo-Nb alloy for the $\{110\}$ plane in bcc with the slabs shifted relative to each other along the $\langle 111\rangle$ direction. Each point corresponds to a symmetrically distinct configurational ordering in the bi-crystal with a glide vector corresponding to the unstable stacking fault. }
    \label{fig:sra_dft}
\end{figure}

The configuration-resolved planar fault energies (CRPF) of 514 symmetrically distinct unstable stacking faults as calculated with DFT is shown in \cref{fig:sra_dft}. The CRPF values vary from $\approx$0.8 J/m$^{2}$ for pure niobium to about 2.0 J/m$^{2}$ in the binary alloy.
The spectrum of values at a particular composition is also found to span a large range of values, suggesting that the state of order among Mo and Nb plays a significant role in determining the CRPF energies.

\begin{figure}
    \centering
    \subfloat[\label{fig:sfe_clex}]{\includegraphics[scale=0.1]{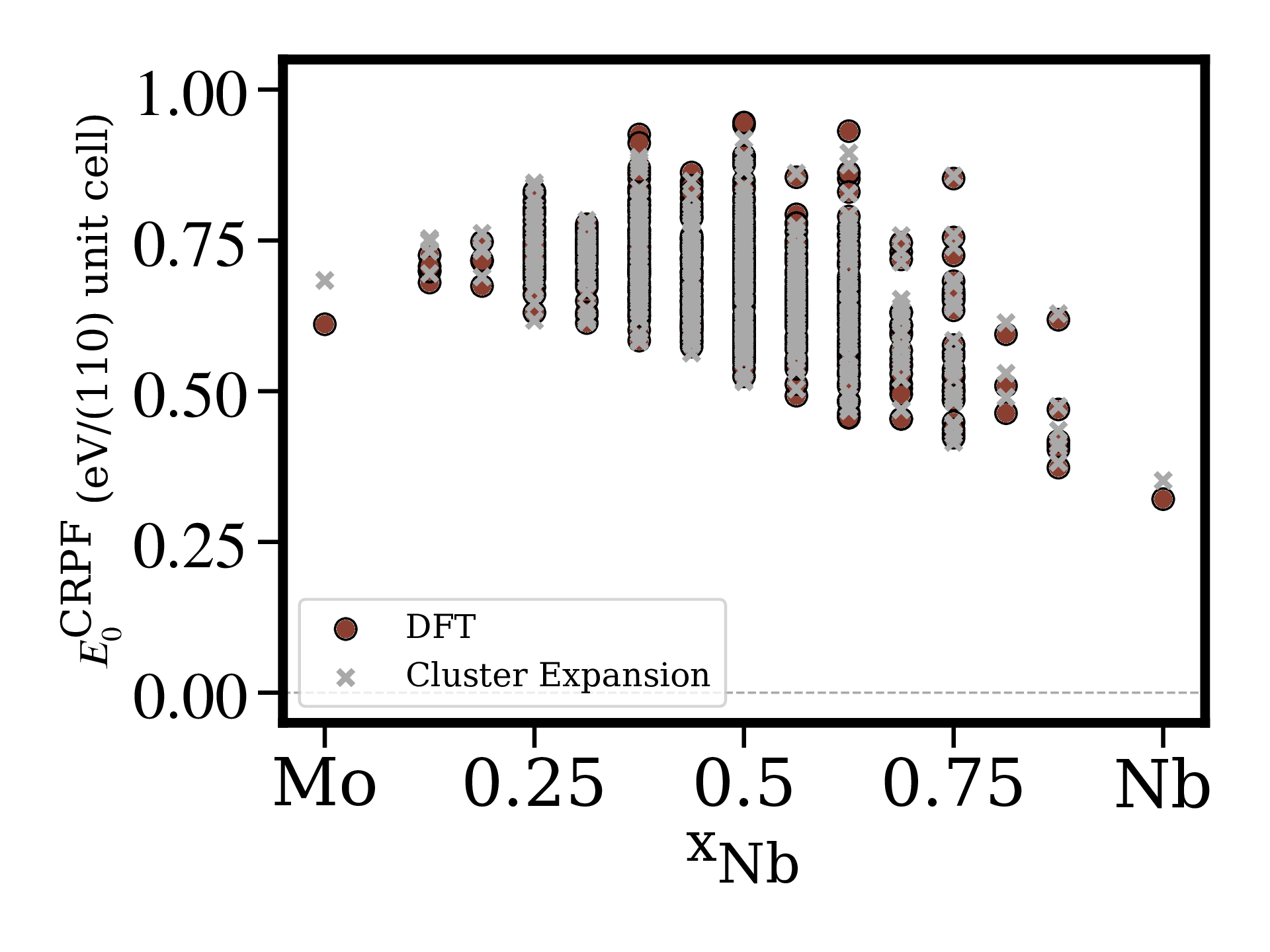}}~
    \subfloat[\label{fig:sfe_clex_validate}]{\includegraphics[scale=0.1]{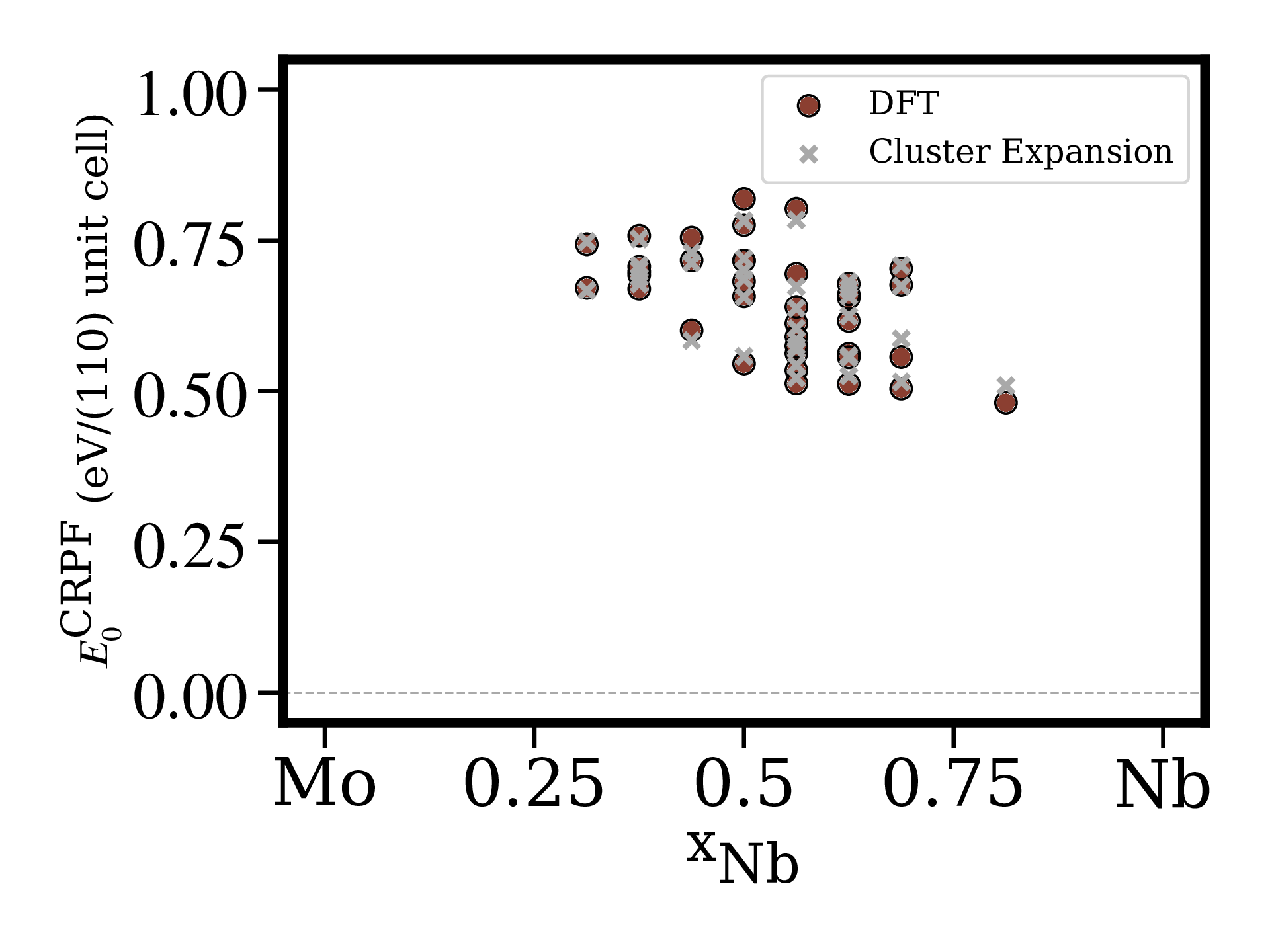}}
    \caption{(a) Comparison of the calculated and cluster expanded \emph{configurationally-resolved planar fault} (CRPF) energies (b) Comparison of the cluster expanded CRPF model against a hold out validation data set calculated with DFT.}
\end{figure}

A cluster expansion was parameterized to describe the dependence of the CRPF energies on the degree of Mo-Nb ordering.
A comparison of the DFT and cluster expanded CRPF energies is shown in \cref{fig:sfe_clex}.
Details about the fitting procedure and cluster expansion model are provided in \cref{sec:methods}.
The unstable stacking fault energies are reproduced well by the cluster expansion model with a fitting error of 0.016 eV per two-dimensional unit cell of the (110) glide plane.
The CRPF energies of configurations that have compositions close to pure molybdenum or niobium have a slightly higher error than configurations with compositions closer to $x=1/2$.
We validated the model by comparing cluster expansion predictions to DFT values of CRPF energies for 38 stochastically enumerated orderings in a 16 atom supercell.
\Cref{fig:sfe_clex_validate} shows a good agreement between the model predictions and the DFT calculations with a validation error of 0.013 eV per unit cell.

Having fit a cluster expansion that accurately describes the unstable stacking fault energy in the binary Mo-Nb system, we next investigated the composition and temperature dependence of this energy. Grand-canonical Monte-Carlo simulations at temperatures above 600K are found to be completely disordered at all compositions, in agreement with experiment\cite{okamoto1991a}. Snapshots of disordered configurations were collected from grand-canonical Monte-Carlo simulations at 600K and 1000K. Chemical potentials were chosen such that the average composition of niobium was 0.25, 0.5 or 0.75. For each Monte-Carlo snapshot, an unstable stacking fault was introduced in the cell and the unstable stacking fault energy was evaluated with \cref{eq:gsf} using the cluster expansions for the CRPF and the formation energy of the bcc Mo-Nb alloy. An unstable stacking fault energy was calculated by introducing a fault in every $(110)$ layer within the simulation cell and for every $\langle 1\overline{1}1\rangle$ direction within the plane. \par

\begin{figure}[htbp]
    \centering
    \subfloat[\label{fig:activation_histogram}]{\includegraphics[scale=0.1]{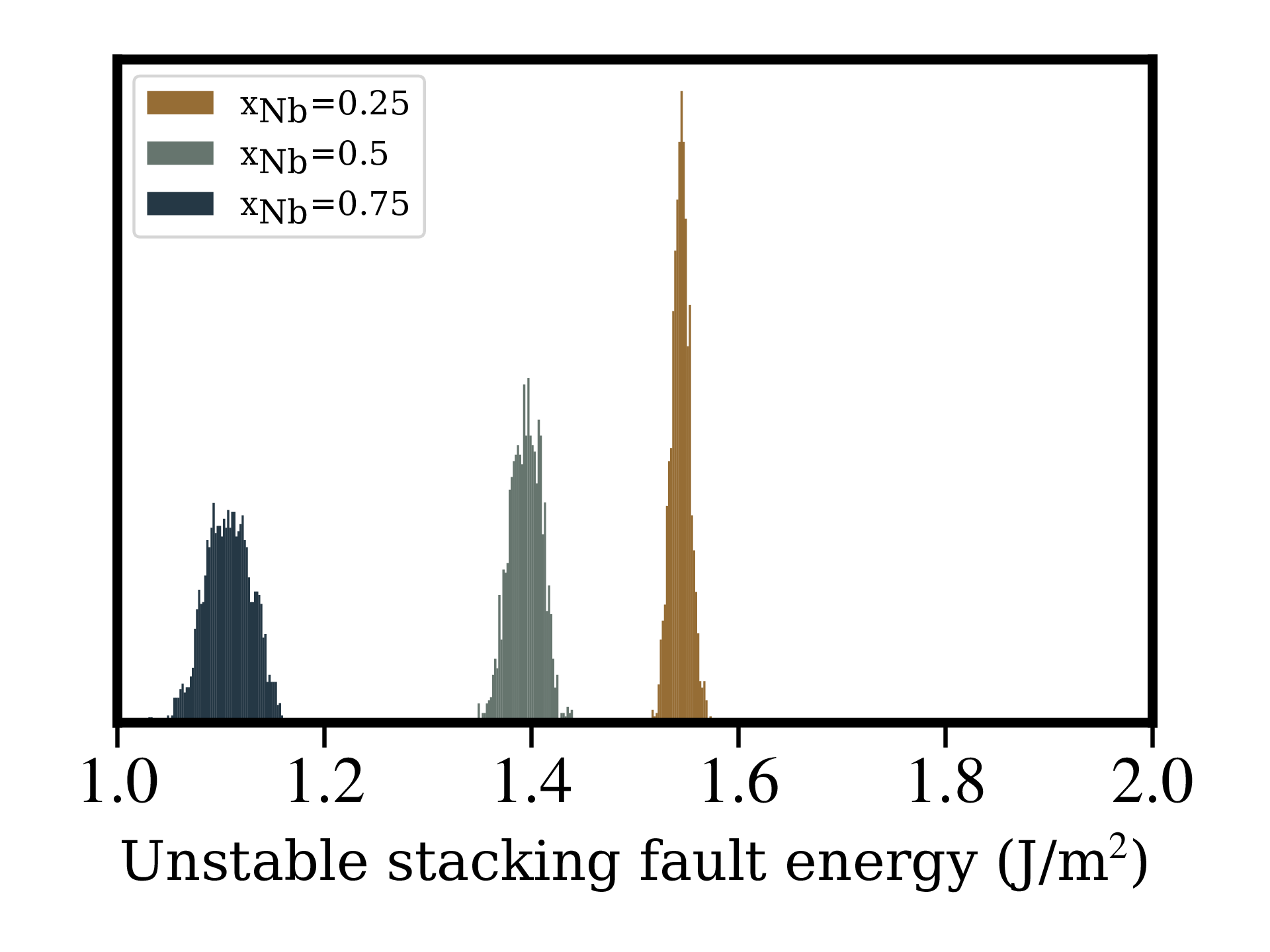}}~
    \subfloat[\label{fig:activation_histogram_temperature}]{\includegraphics[scale=0.1]{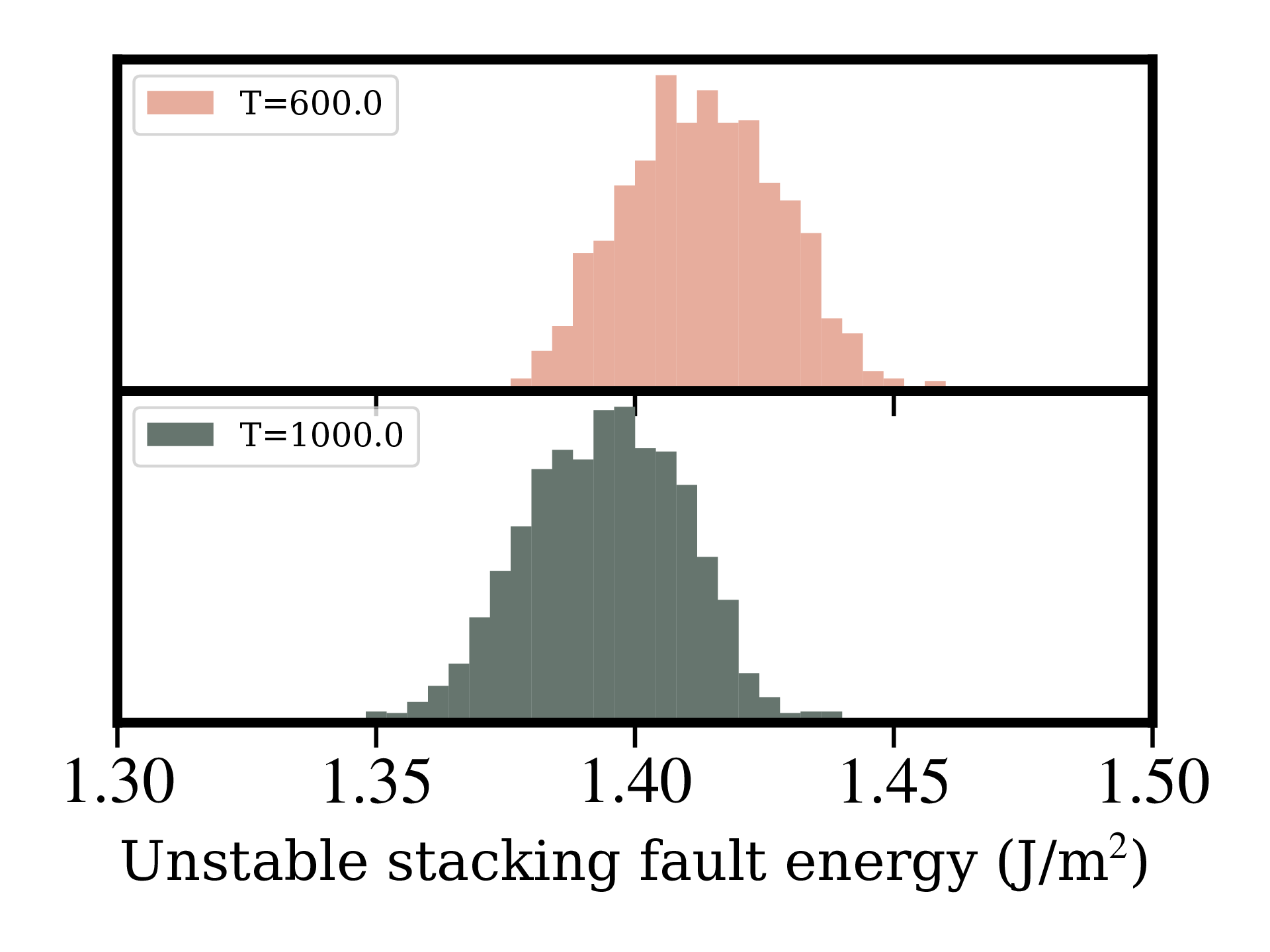}}
    \caption{(a) The composition dependence of the unstable stacking fault energies in the binary Mo-Nb alloy at 1000K. The average unstable stacking fault energies are collected in snapshots from grand-canonical Monte-Carlo calculations at appropriate chemical potentials. (b) The temperature dependence of the unstable stacking fault energies at 600 and 1000K at an average simulation cell composition of 0.5}
\end{figure}

\Cref{fig:activation_histogram} shows a histogram of unstable stacking fault energies at three different niobium compositions at a temperature of 1000K. The unstable stacking fault energies decrease with increasing niobium composition. This is in agreement with the general trend of CRPF energies across compositions in \cref{fig:sra_dft}. Our results predict that the stacking fault energies  vary strongly with the average composition of the alloy. The spectrum of USF energies at elevated temperatures are normally distributed. The distribution is very sharply peaked at a niobium composition of 0.25 with a slightly more broadened distribution with increasing niobium composition. \par

We investigate the temperature dependence of the unstable stacking fault energies in \cref{fig:activation_histogram_temperature}. Grand-canonical Monte-Carlo calculations were performed at temperatures of 600K, and 1000K at chemical potentials that corresponded to average alloy compositions of 0.5. \Cref{fig:activation_histogram_temperature} shows that the distribution of unstable stacking fault energies continues to be normally distributed across a wide range of temperatures. The magnitude of the fault energy decreases with increasing temperature. The distribution of energies is also broadened with increasing temperature. These results suggest that the average unstable stacking fault energies vary strongly with niobium composition, while the distribution of energy values is sensitive to the temperature.

\section{Discussion and Conclusion}
\begin{figure}[htbp]
    \centering
    \includegraphics[scale=0.1]{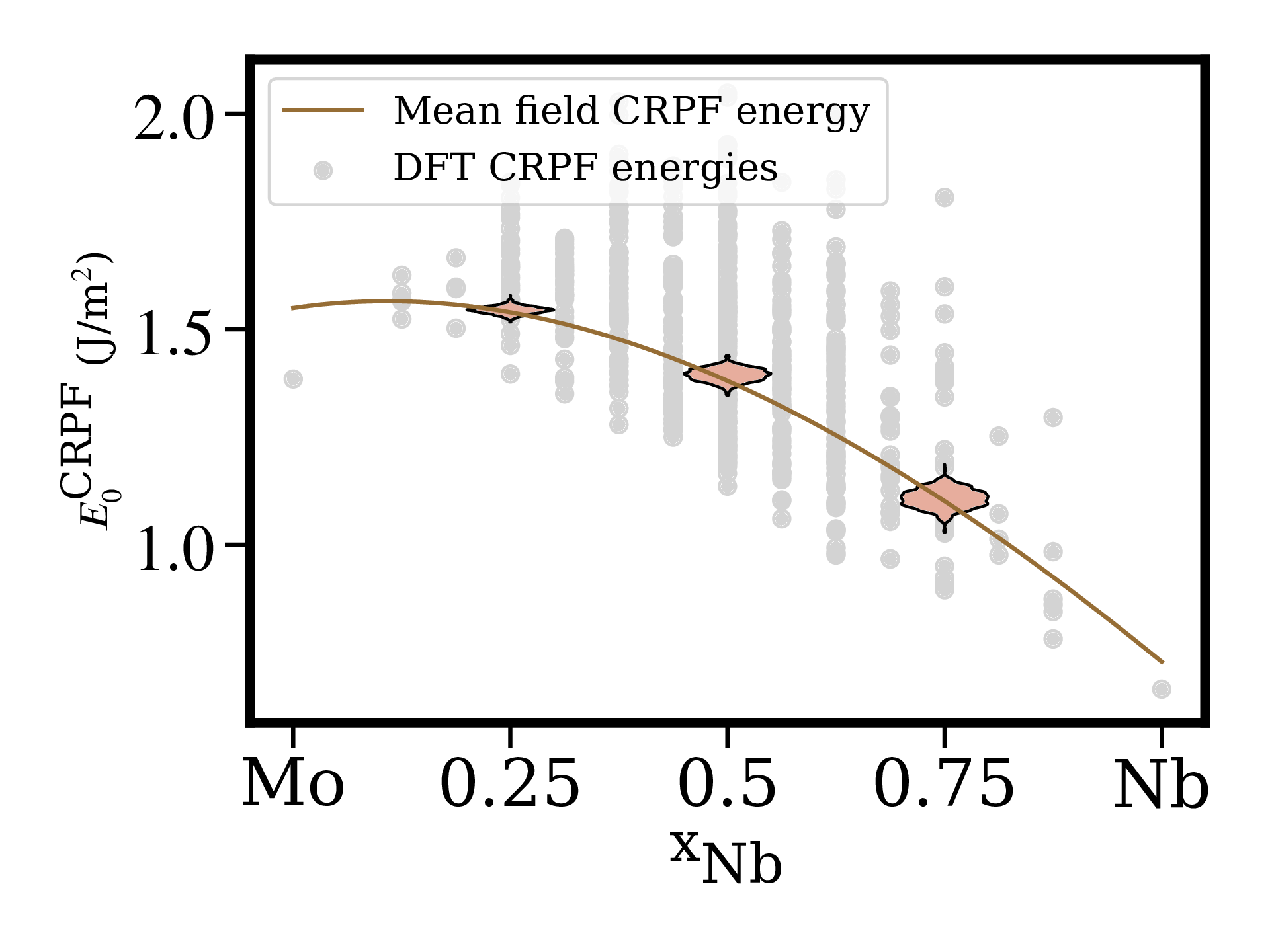}
    \caption{The figure shows the DFT calculated CRPF energies compared to the energies from a mean-field approximation and the distribution of values from Grand-Canonical Monte-Carlo simulations at 1000K. The distributions from Monte-Carlo simulations are shown as violin plots with the breadth of the plot at a given niobium composition corresponding to the relative number of configurations with that CRPF energy.}
        \label{fig:mean_field_mc}
\end{figure}

We have developed a rigorous approach to describe the dependence of generalized stacking fault energies on the degree of ordering in multicomponent alloys. The approach relies on the decomposition of the bicrystal energy into a long-range configurational contribution and a local planar fault energy, referred to as a \emph{configurationally-resolved planar fault energy} (CRPF). The dependence of the CRPF on configuration is then represented with a short range cluster expansion over sites within a chemical interaction range of the glide plane, while the long-range configurational contribution is captured with the conventional cluster expansion as originally introduced by Sanchez et al \cite{sanchez1984}. The formalism is applied to quantify the unstable stacking fault energy in a prototypical binary Mo-Nb alloy. Monte-Carlo calculations informed by accurate configurational and CRPF cluster expansions predict a strong composition and temperature dependence for the average USF energy. \par

The tools we have developed here enable rigorous statistical mechanics studies of the effects of short and long-range order on the generalized stacking fault energies in multicomponent alloys.
For example, \cref{fig:mean_field_mc} compares the DFT CRPF energies, the predicted CRPF distributions from Monte-Carlo simulations at 1000K, and the CRPF energy as a function of composition for a fully disordered random solid solution in the Mo-Nb binary.
The spread of the sampled CRPF energies at elevated temperatures in disordered alloys is much smaller than the full range of values that may exist across all symmetrically distinct arrangements of Mo and Nb atoms.
For instance, at a composition of $x_{\textrm{Nb}}=0.5$, CRPF values of distinct binary orderings vary by almost a factor of two between ~1-2 J/m$^{2}$.
In contrast, the Monte-Carlo simulations at the same composition predict a distribution that is sharply peaked around 1.4 J/m$^{2}$, with a spread of only 0.05 J/m$^{2}$. This suggests that properties of disordered alloys may be difficult to extract directly from the spectrum of calculated CRPF energies without a statistical mechanics treatment. \par

\Cref{fig:mean_field_mc} also shows that the CRPF energies sampled in the Monte Carlo simulations at elevated temperatures and a mean-field estimate are very similar.
The mean-field estimate of \cref{fig:mean_field_mc} was calculated by substituting the correlations of a random alloy in the cluster expansion of the CRPF.
The mean-field approximation, therefore, neglects any long or short-range order that may exist in the actual alloy and in the Monte Carlo simulations.
The fact that the mean-field estimate is very close in magnitude to the Monte-Carlo averages suggests that short-range order does not play a significant role in affecting the unstable stacking fault in the Mo-Nb alloy at 1000 K.
The spread of CRPF values, however, varies strongly with niobium composition, suggesting that niobium rich environments have a slightly broader distribution of energies than molybdenum rich environments.\par

The formalism presented in this study brings us closer to a truly rigorous multi-scale model of dislocation motion and evolution in multicomponent alloys.
In conjunction with techniques to estimate the effect of configurational disorder on transport coefficients\cite{vanderven2005,deng2015,vanderven2010}, structural phase transitions\cite{thomas2013}, surface and bulk thermodynamics\cite{natarajan2017c,vanderven2004,thomas2010}, this formalism expands our palette of models to naturally account for the mechanical behavior of engineering materials.
 Phase-field models of dislocation motion can describe the motion, formation and structure of dislocations in a variety of crystal structures and material systems.
 As we move into complex multicomponent alloy chemistries, formalisms that can estimate the GSF energies across vast composition spaces and elevated temperatures are crucial to establishing a rigorous links between electronic structure calculations and phenomenological theories of mechanical properties of materials.

\section{Methods}
\label{sec:methods}
\begin{figure}
    \centering
    \includegraphics[scale=0.1]{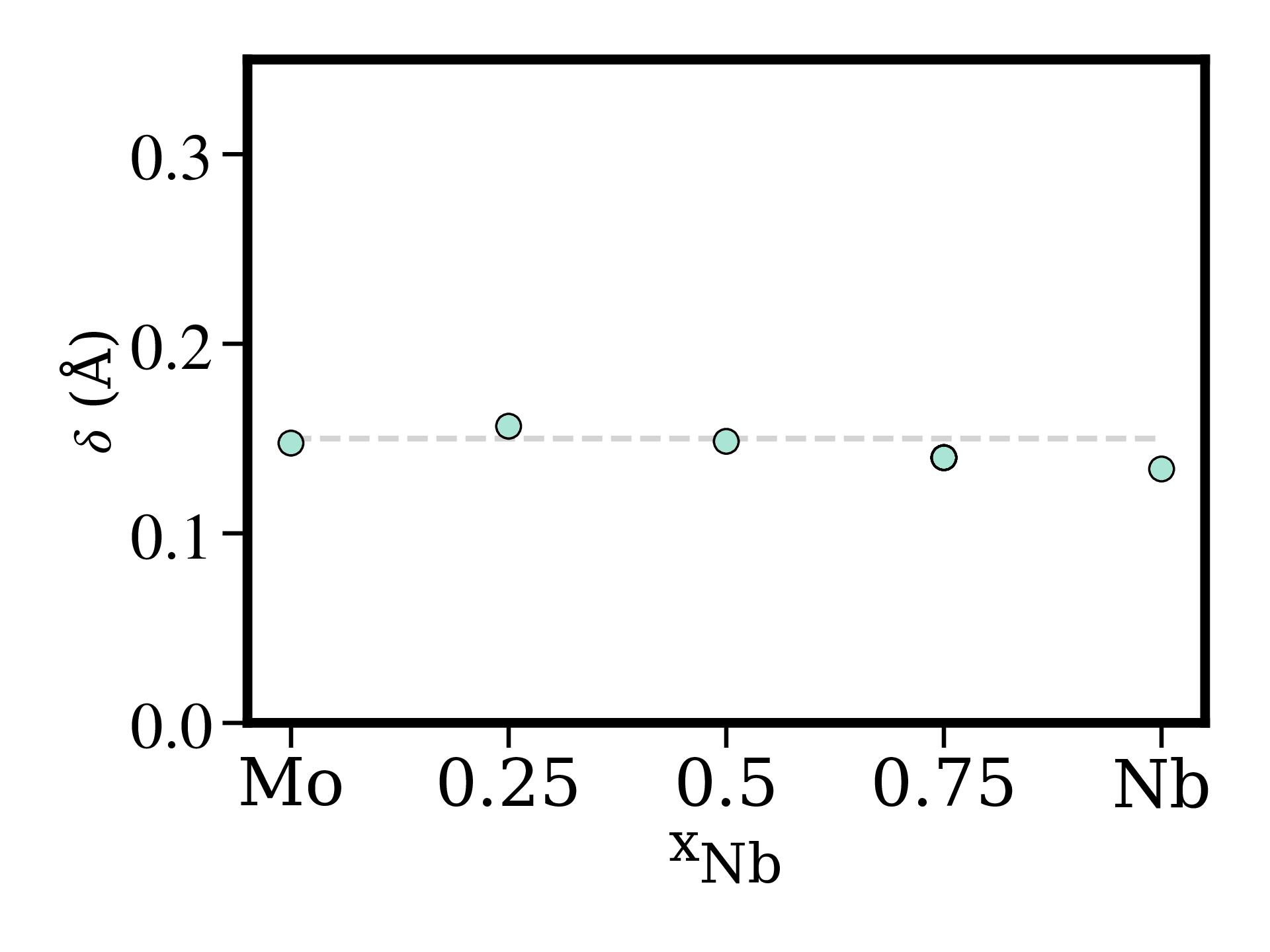}
    \caption{The figure shows the separation distances ($\delta$) at the minimum energy for the unstable stacking fault of the ordered phases in \cref{fig:sfe_ordered_phases}}
    \label{fig:delta_ordered_phases}
\end{figure}
Total energies of crystals were calculated within the generalized gradient approximation to desity functional theory as parameterized by Perdew-Burke-Ernzerhof \cite{perdew1996} and implemented in the \emph{Vienna Ab-Initio Simulation Package}\cite{kresse1996a}.
The projector augmented wave (PAW) method was used to describe the interaction of valence electrons with core states.
The PAW potentials treated the semi-core $s$ electrons as valence states.
The planewave cutoff was set to 480 eV and an automatic k-point grid with 42 k-points \AA$^{-1}$ were used for Brillouin zone integration. The total energies of configurations in a structure corresponding to the unstable stacking fault along the $[1\overline{1}1]$ direction within the $(110)$ plane was calculated with a static calculation. The separation distance between the two crystal halves was set to 0.15\AA at the pure Mo composition and all lattice parameters were homogeneously scaled based on Vegards law\cite{vegard1921}. The composition dependence of the lattice parameters are informed from the benchmark calculations shown in \cref{fig:monb_vol}. The equilibrium separation $\delta$ was found to vary with composition, however a separation scaled by the volume of the crystal relative to that of pure Mo defined as $\tilde{\delta} = \delta / (\frac{V}{V_{Mo}})^{\frac{1}{3}}$ was found to be independent of composition, as shown in \cref{fig:delta_ordered_phases}. Initial GSF energy calculations were performed with the \texttt{multishifter} code\cite{goiri2019}.\par

Cluster expansion Hamiltonians were parameterized with the \emph{Clusters Approach to Statistical Mechanics} (CASM) software package\cite{vanderven2018,puchala2013,thomas2013,casm}.
Symmetrically distinct configurations were generated on the parent bcc and unstable stacking fault structure with the CASM software package. All configurations in symmetrically distinct supercells containing up to six atoms were enumerated in bcc. All configurations were enumerated in the primitive structure containing the unstable stacking fault with 8 $(110)$ layers. 377 stochastic orderings were also generated in cells containing 16 atoms. Two separate cluster expansions were subsequently trained. Clusters on the bcc crystal structure were chosen from a pool of pairs, triplets and quadruplets with a maximum length of 10.3,8.3,7.3 \AA with the genetic algorithm informed with a 10-fold cross-validation score. The resulting RMSE was 0.0008 eV/atom with a cross-validation score of 0.0008 eV/atom. The clusters for the unstable stacking fault energy included pairs and triplets with a maximum length of 7.3, and 5.3 \AA. DFT calculations were used to parameterize the cluster expansion Hamiltonian with the L$_{1}$ regularization to least squares. The regularization parameter was chosen with a 10-fold cross-validation metric. \par

Grand-canonical Monte-Carlo calculations were performed with a simulation cell containing 16 $[110]$ layers and a total of 1600 atoms across a range of chemical potentials and temperatures. Configurational snapshots were extracted from the Monte-Carlo simulations every 10 passes after the system was determined to be equilibrated. Within each snapshot, the CRPF was calculated with the CRPF cluster expansion by introducing a planar fault within a $(110)$ layer. The unstable stacking fault energies were recorded for each $(110)$ plane in the simulation cell for a planar fault obtained by translating the two crystal halves along every $\langle111\rangle$ direction in the plane. The stacking fault energies were subsequntly calculated by dividing the activation energy by the area of the $(110)$ plane in a simulation cell with a lattice parameter scaled in accordance with Vegards Law.

\section{Acknowledgements}
We are grateful for financial support from the ONR BRC Program, Grant Number N00014-18-1-2392.
Computing resources were provided by the National Energy Research Scientific Computing Center (NERSC), a U.S. Department of Energy Office of Science User Facility operated under Contract No. DE-AC02-05CH11231 and the Center for Scientific Computing (CSC) supported by the California NanoSystems Institute and the Materials Research Science and Engineering Center (MRSEC; NSF DMR 1720256) at UC Santa Barbara with funds from the National Science Foundation (CNS-1725797).

\end{document}